\documentclass[epj]{svjour}

\usepackage{graphicx}
\usepackage{fancyhdr}
\def\makeheadbox{{
 }}
\def\mytitle{My title} 
\def\myauthors{My name}  
\def\mytype{My type of session}
\def\mysession{My session}

\def\mytitle{Uncertainties of the cosmic antiproton flux} 
\def\myauthors{Iris Gebauer}    
\def\mytype{Contributed Talk}    
\def\mysession{Cosmology and Astrophysics}

\begin{document}
\title{Uncertainties of the antiproton flux from Dark Matter
annihilation in comparison to the EGRET excess of diffuse gamma rays}
\author{Iris Gebauer\inst{1}
\thanks{\emph{Email:} gebauer@ekp.uni-karlsruhe.de}%
}                     
%
%
\institute{Institut f\"{u}r Experimentelle
Kernphysik, Universit\"{a}t Karlsruhe (TH), P.O Box 6980, 76128 Karlsruhe, Germany}
%
\date{}
\abstract{
The EGRET excess of diffuse Galactic gamma rays  shows all the features
expected from dark matter annihilation (DMA): a spectral shape
given by the fragmentation of mono-energetic quarks, which is the same in all sky
directions and an intensity distribution of the excess expected
from a standard dark matter halo, predicted by the rotation curve.
From the EGRET excess one can predict the flux of antiprotons
from DMA. However, how many antiprotons arrive at the detector strongly depends on the pro\-pagation model. The conventional isotropic propagation
models trap the antiprotons in the Galaxy leading to a local
antiproton flux far above the observed flux. According to Bergstr\"{o}m
et. al. this
excludes the DMA interpretation of the EGRET excess. Here it is shown that more
realistic anisotropic propagation
models, in which most antiprotons escape by fast transport in the z-direction, are consistent
with the B/C ratio, the antiproton flux and the EGRET excess from DMA.
} 
\maketitle
%

\section{Introduction}
\label{intro}
The interpretation of the observed EGRET excess of diffuse Galactic
gamma rays as Dark Matter annihilation (DMA) (see \cite{us} or
contributions by W. de Boer, C. Sander and
M. Weber, this volume) could be a
first hint at the nature of dark matter. 
The excess was observed in all sky directions. From the spectral shape of the excess the
WIMP mass was constrained to be between 50 and 100 GeV and from the distribution of
the excess in the sky the Dark Matter (DM) halo profile was obtained. One of the most important
criticisms of this analysis was a paper by Bergstr\"{o}m et. al. \cite{bergstrom1} claiming that the antiproton flux from
DMA would be an order of
magnitude higher than the observed antiproton flux. 
They used a conventional propagation
model assuming the propagation of charged particles to be the same in the halo and
the disk.
However, the propagation in the halo (perpendicular to the disk) can be
much faster than the propagation in the disk \cite{breit}.

In this paper we show that the
local antiproton flux from DMA can be strongly reduced in an anisotropic
propagation model and that the DMA interpretation of the EGRET excess
can by no means be
excluded by Galactic antiprotons.
In section \ref{CM} we discuss the pro\-blems of the isotropic model for cosmic
ray transport leading to the fact that our galaxy can work as a large
storage box for antiprotons. An anisotropic pro\-pagation model,
which simultaneously describes the \\
EGRET excess and and the observed
local fluxes of charged cosmic rays is introduced in section
\ref{CP}. Section \ref{conclusion} summarizes the results. 

\section{Antiproton flux in an isotropic propagation model\label{CM}}
\begin{figure*}
\begin{center}
 \includegraphics [width=0.45\textwidth,clip]{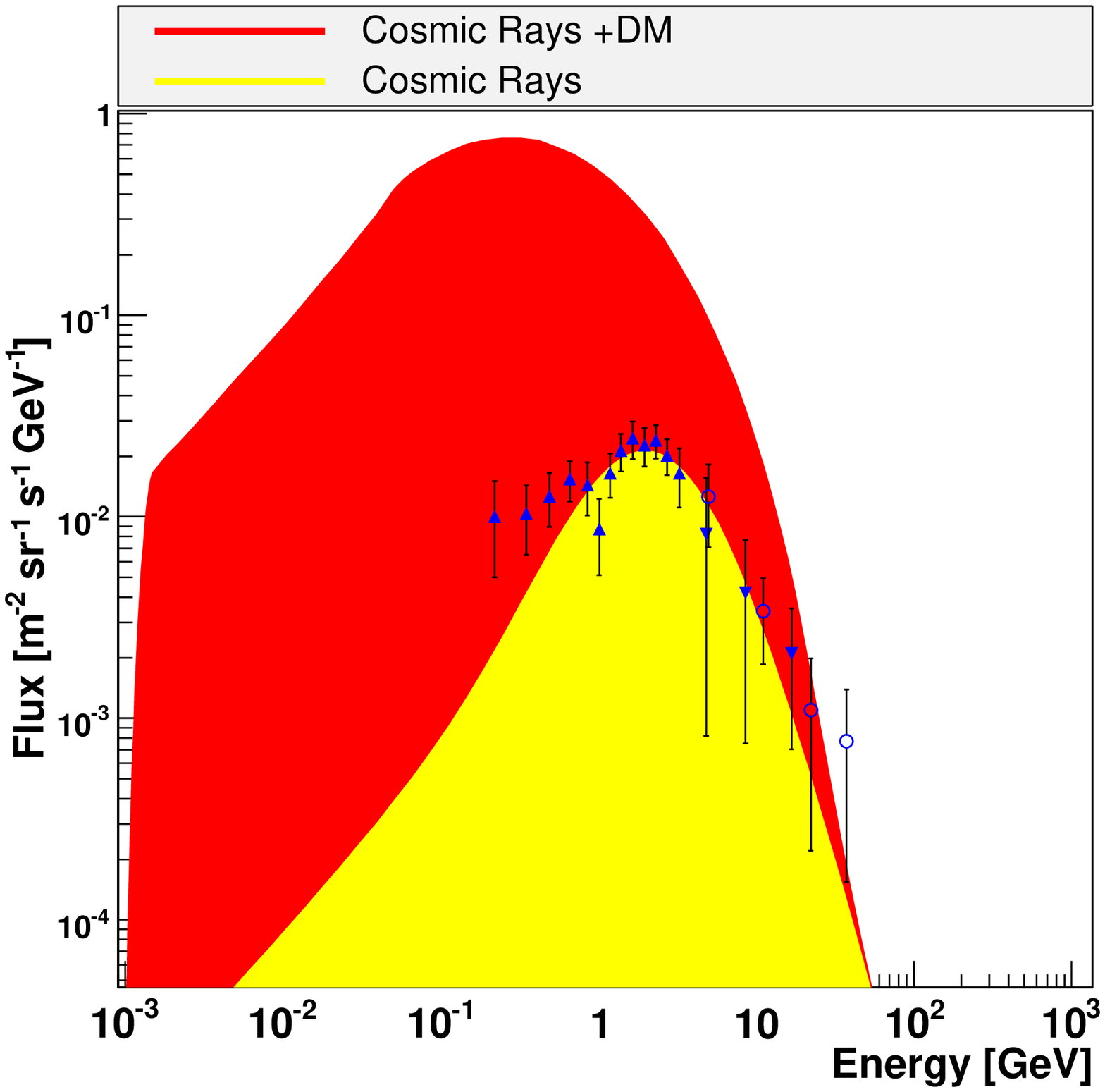}
\includegraphics [width=0.45\textwidth,clip]{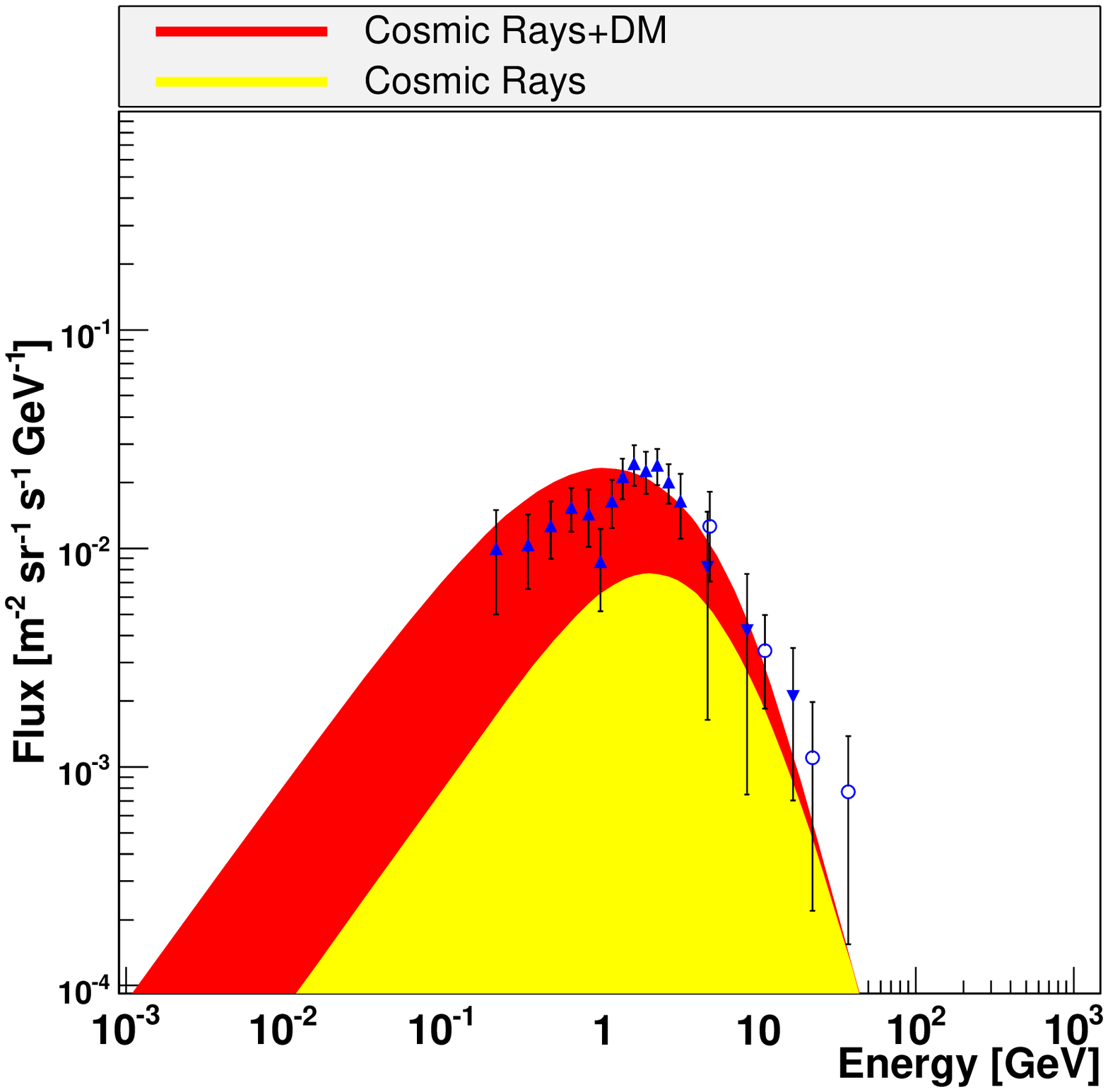}
 \caption[]{Comparison of the antiproton production including DMA
  in a model with isotropic propagation (a)
 and a model with anisotropic propagation and trapping between
 molecular clouds (b).}
 \label{fig1}
\end{center}
\end{figure*}
In order to explain the observed isotropy of Cosmic Ray (CR) fluxes one assumes the CRs
to perform a random walk in all directions by scattering on randomly oriented
turbulent magnetic fields inside the  plasma. In this case the propagation is governed by a diffusion equation, which
can be solved for the steady state case numerically.
From the ratio of unstable/stable nuclei (like $^{10}Be/^{9}Be$) one obtains the average residence time of CRs in the Galaxy to be of the order of $10^7$ yrs.
Particles can be lost either by fragmentation, decay or just leaving the
Galaxy to outer space. Since they travel with re\-lativistic speed the
long residence time requires that they cannot move rectilinear to outer
space, but must be scattering many times without loosing too much
energy. During their journey CRs may interact with the gas in the Galaxy
and produce secondary particles. The ratio of secondary/pri\-mary
particles, like the B/C ratio is a measure for the amount of traversed matter (grammage) by CRs
during their lifetime $t_{CR}$. The grammage is given by $\rho c t_{CR}$, where
$ct_{CR}$ is the path length for a particle traveling with the speed of light $c$.
It was found to be of the order of $10 g/cm^{2}$ \cite{schlickeiser,berezinsky},
which corresponds to a density of about 0.2 $atoms/cm^{3}$. This is significantly
lower than the averaged density of the disk of 1 $atom/cm^{3}$, which suggests
that CRs travel a large time in low density regions. In an isotropic
propagation model this would be the thin halo. However, as we will see
in section \ref{CP} this interpretation is strongly model dependent.\\
An excellent program providing a numerical solution to the diffusion
equation for CRs
is the publicly available GALPROP code \cite{galprop,galprop1}.  
The standard model for isotropic CR transport in GALPROP does not explain the EGRET excess of
diffuse gamma rays. This could be remedied partially by applying strong breaks to the injection spectra
of protons and electrons in order to obtain a higher intensity of protons and
electrons above a few GeV, but the intensity below this break had to stay the same
in order not to modify the gamma ray spectrum below 1 GeV \cite{optimized}. These
breaks are only applied for the electrons and protons, not for the other nuclei in
order not to upset the B/C ratio. Different breaks in the injection spectra for
different nuclei implies different acceleration histories for different nuclei. In
addition, assuming the locally observed spectrum to be different from the spectra
elsewhere in the Galaxy is unexpected, since  diffusion is fast compared
to the
energy loss time, so diffusion equalizes the spectrum everywhere in agreement with
the observation that the gamma ray spectra in all directions can be
described by the {\it same} CR spectrum. If
one attributes the EGRET excess to a new source, like DMA, one runs into the
problem of a too large flux of antiprotons, as discussed in detail in \cite{bergstrom1}. We have implemented the DMA as a source term into the publicly
available GALPROP code \cite{galprop,galprop1} and find a similar result, as shown in
Fig. \ref{fig1}a.  This is not surprising, since GALPROP uses the same priors as
the program used by \cite{bergstrom1}: (i) the propagation is dominated by diffuse
scattering, which is assumed to be the same in the halo and the
disk (ii) the gas in the disk is smoothly distributed (iii) the
influence of the observed static magnetic fields can be neglected. The main
reason for the large flux of antiprotons from DMA in such a model is
the long residence time of charged
particles ($10^7$ yrs), which requires all particles to spend most of their
lifetime in the thin galactic halo and enter and exit the dense galactic disk
multiple times thus acquiring grammage. In this case antiprotons from
DMA are trapped in the Galactic halo, just like conventional CRs, and DMA increases the averaged density of
antiprotons by  orders of magnitude, so the flux of antiprotons becomes of the same
order of magnitude as the EGRET excess. Note that the production ratio of
antiprotons/gammas from DMA is only at the percent level, as is well known from
accelerator experiments for the fragmentation of mono-energetic quarks,
so the enhancement of antiprotons comes from the propagation model, not
from the production.


\section{Antiproton fluxes in an anisotropic propagation model}\label{CP}

\begin{figure*}
\begin{center}
 \includegraphics [width=0.45\textwidth,clip]{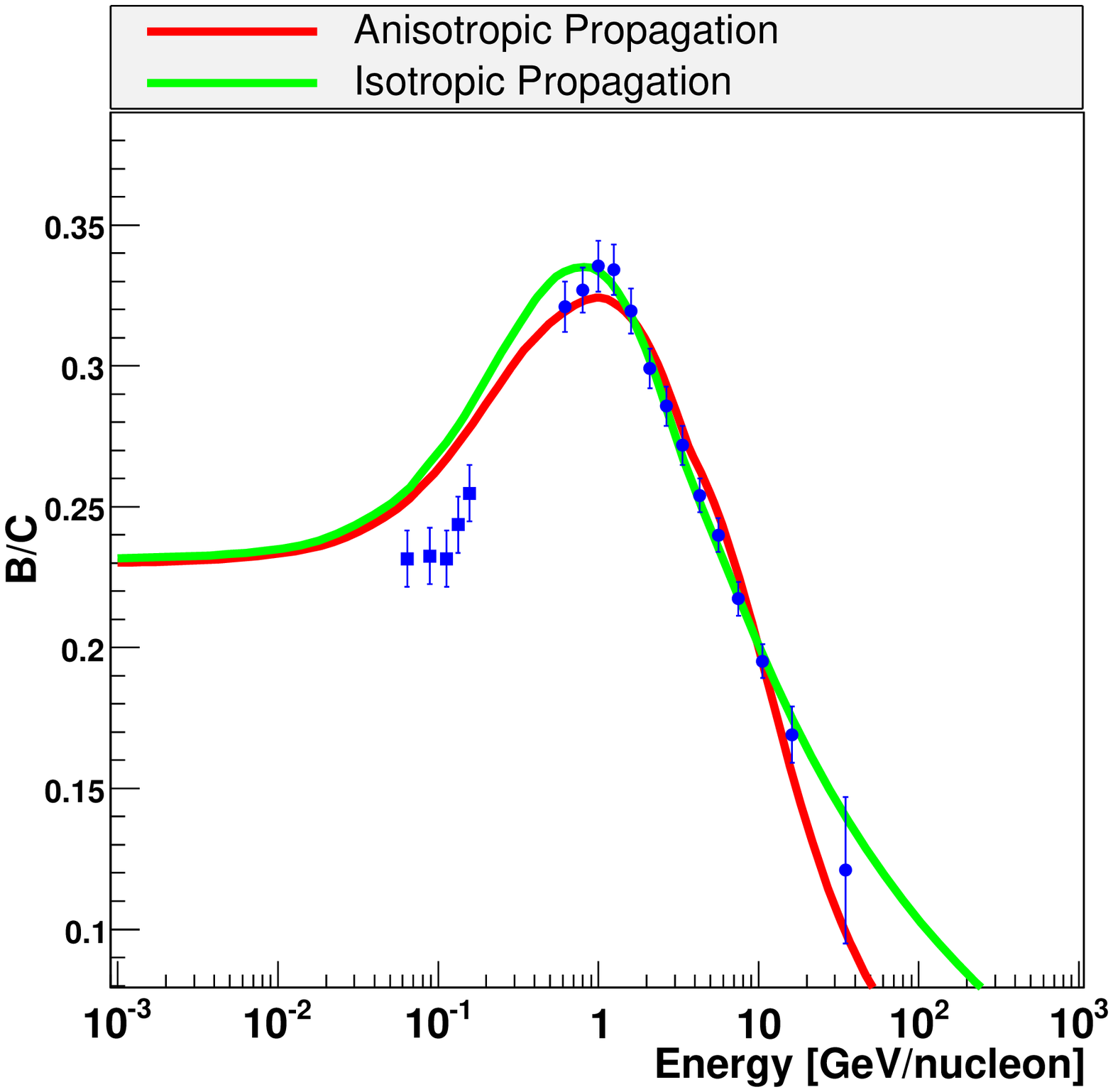}
\includegraphics [width=0.45\textwidth,clip]{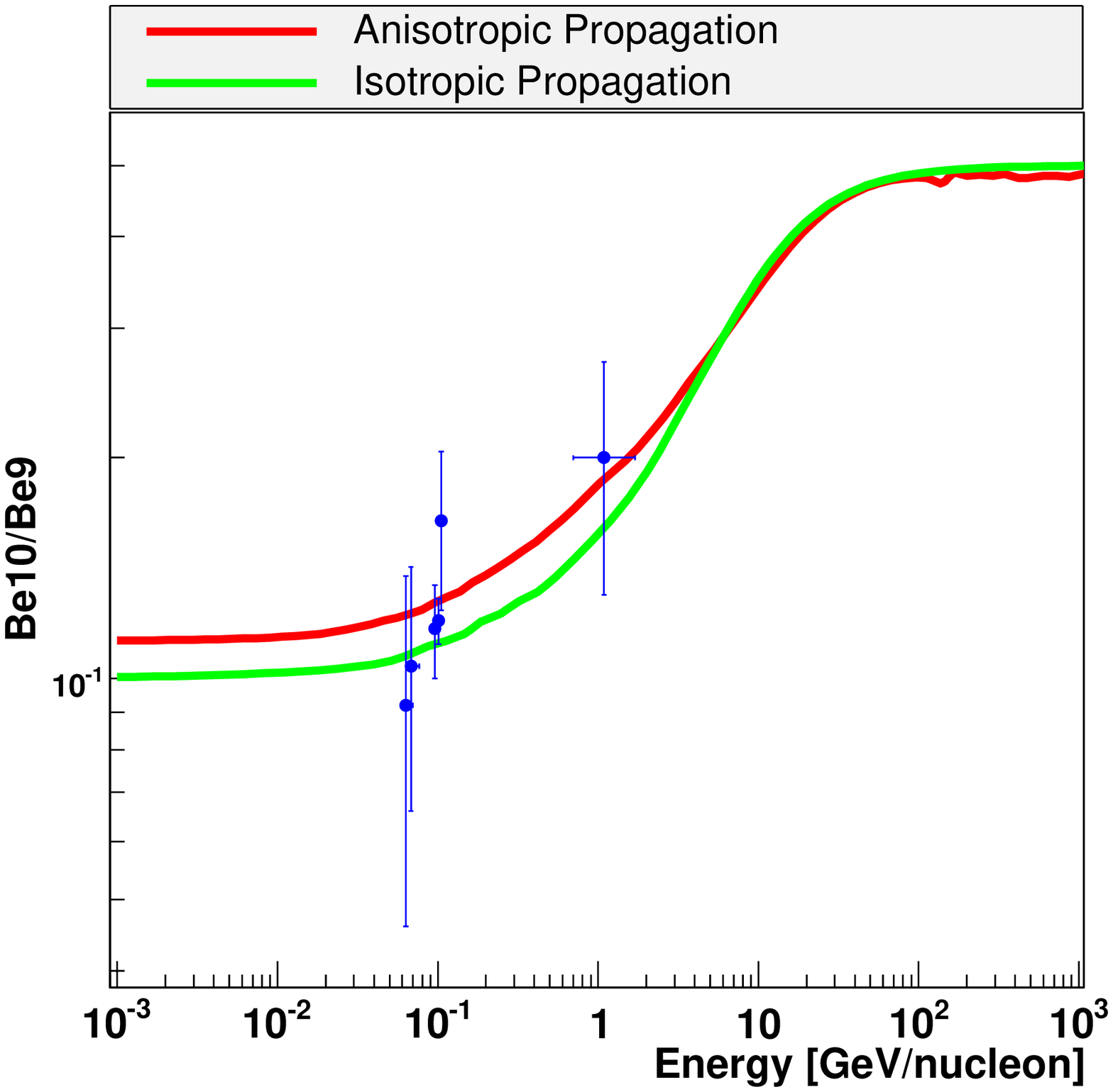}
 \caption[]{The B/C ration (a)and the beryllium-fraction (b) for
 an anisotropic diffusion model with trapping between MCs (red) and for a
 model with isotropic diffusion (green).}
 \label{fig2}
\end{center}
\end{figure*}

The  propagation picture  with isotropic propagation is based on hydromagnetic wave
theories, in which the random (small-scale) component of the magnetic fields
dominates over or are of the same order of magnitude as the regular large scale
components.
\begin{table}
  \begin{center}
    \begin{tabular}{|l|l|l|}
      \hline
      Name& Symbol &  value  \\
      \hline
Diffusion in x, y&$D_{xx}, D_{yy}$ & $5.8 \cdot 10^{28} cm^2/s$ \\
Diffusion in z&$D_{zz}$ & $3.0 \cdot 10^{30} cm^2/s$\\
break rigidity &$\rho_{0}$& $4.0 \cdot 10^{5} MV $\\
energy dependence& $\alpha$ for $\rho <\rho_{0}$&$0.33$ \\
 & $\alpha$ for $\rho >\rho_{0}$&$0.6$\\
Convection &$V_{0}$ & $250 km/s$\\
Convection slope&${dV \over dz}$& $37 km/s/kpc$\\
grammage paramater &$c$& $12$\\
      \hline
    \end{tabular}
  \end{center}
  \caption{Parameters of an anisotropic propagation model.The components
 off the diffusion tensor are given by $d_{ii}(\rho)=\beta \cdot D_{ii} ({\rho \over \rho_{0}})^{\alpha}$, convection is given by $V(z)=V_{0} \cdot \theta(z-0.1kpc)+{dV \over dz} z$. Note that this
 set of parameters is not unique, since they are all correlated.} \label{tab1}
\end{table}
From the isotropy of the CRs one assumes that the regular components of the magnetic
fields can be neglected, so there is no preferred direction for Alfv\'en waves.
The turbulent component is locally as large as 10 $\mu G$, while the
regular field is only about 3 $\mu G$ \cite{heiles,han0}. However, even
if the turbulent small scale and regular large scale components are of
the same order of magnitude the
ratio of perpendicular/parallel diffusion is  about 0.1 (see
\cite{breit} and references therein), which
implies that the CRs still preferentially follow the regular magnetic
field lines, as demonstrated by following the trajectories of CRs in models of the Galactic
magnetic fields \cite{blasi,codino}.
The regular component has strongly preferred directions: in the disk it is toroidal
with a maximum at about 150 pc above and below the disk and with an additional poloidal field with
its maximum in the centre of the Galaxy.
For fast parallel diffusion this implies that CRs preferentially move along the spiral fields just
above or below the disk or they follow the polodial component into the halo. 
An additional effect concerning charged particles may be related to mole\-cular
clouds: the gas density in the disk varies from $10^{-3}/cm^{3}$ in the warm
ionized medium to $10^2/cm^{3}$-$10^3/cm^{3}$  in clumps of cold gas with a size
of a few pc. In the center the density may be as high as $10^7/cm^{3}$ in dense molecular clouds (MCs), where star formation occurs. On average
the gas density is $1/cm^{3}$ in the disk. Inside MCs magnetic fields far
above the random components have been observed (see \cite{heiles} for a
review). What is more important, these fields seem to be correlated with the
observed static magnetic fields outside the MCs \cite{han}.  This can only be
understood, if the MCs remember the large scale magnetic fields in the interstellar
medium, i.e. if during the contraction flux freezing occurs. In this case the
magnetic field lines from the ISM will become highly concentrated near
the MCc and the MCs will form a network of interconnected clouds,
focussing the magnetic field lines towards them.
CRs in the ISM following these field lines will be reflected by the  concentration
of the field lines. As worked out by Chandran \cite{chandran}, the MCs
can act as magnetic
mirrors for CRs, just like the concentration of magnetic field lines near the poles
from the earth trap the CRs in the famous Van Allen radiation belts. The large distances (pc
scale) between the MCs allows to trap particles up to the TeV scale, thus increa\-sing
the grammage and the residence time. In such a setup particles acquire
grammage and age in the low density regions in the disk (between MCs) and
not in the halo as in the isotropic propagation model. 
Thus, the halo size is not a sensitive parameter anymore and particles,
once in the halo, will be preferentially transported away from the disk
by a combination of convection and fast diffusion along the regular poloidal field lines in the halo.
We have implemented this propagation picture in the
publicly available source code of GALPROP by (i) allowing for a
diffusion tensor instead of a diffusion constant; (ii)
allowing an inhomogeneous grid in order to have step sizes below 100 pc in the disk region and large step sizes in the
halo; (iii) implementing the dark matter annihilation  as a source term of
stable primary particles, especially antiprotons, positrons and gamma
rays, in the diffusion equation. The dark matter distribution was taken
to be the one obtained from the EGRET excess \cite{us}. The
grammage and escape time were adjusted for charged particles to account
for the fact that secondary particles are now produced largely locally,
since particles produced far away from the solar system are likely to
diffuse into the halo, thus escaping to outer space.  If the trapping in
MCs is effective, one would expect it to increase the grammage and
residence time by the same factor, called grammage parameter c. Since the trapping mechanism is
independent of energy and only depends on the pitch angle the
trapping can be modeled by a constant c. The most important GALPROP
parameters have been summarized in Table \ref{tab1}. The transport from
the disk to the halo is quite uncertain, since the magnetic field lines have to be continuous, which
implies they must connect from the toroidal field to the halo. It should
be noted that the average scale height of SNIa is expected to be about
350 pc (thick disk) and  the ejecta connect to the halo in chimney like
structures (see e.g. \cite{breitschwerdt} and references therein), which
can drive magnetic field lines towards high altitudes ( $\approx$ 1
kpc), thus facilitating the transport to the halo by the fast parallel
diffusion. This was simulated as an enhanced convection term starting at $V_0=250 km/s$ at 100 pc above the disk and then increasing with
the distance $z$ above the disk as $dV/dz=37 km/s/kpc$. It should be mentioned that this set of parameters is not unique,
since they are all correlated. As shown in Figs. \ref{fig1}b
and \ref{fig2} the B/C ratio, the $^{10}Be/^9Be$ ratio and the
antiproton flux are all well described by this set of
parameters. Note that
most of the antiprotons from DMA in the halo diffuse into outer space and are never
observed in the detector in contrast to the diffusion model with isotropic
diffusion (Fig. \ref{fig1}a). It should also be noted that the antiprotons produced
in the ringlike structures of DM near us are unlikely to reach us, since they
follow the toroidal magnetic field lines, so they diffuse fast in the $\phi$
coordinate, not in $r$. Such a propagation would require tuning the 3D version of
GALPROP. However, this takes an excessive amount of CPU time and memory
and will be subject of future studies.
It should be kept in mind that the convection speed and the diffusion
coefficient in the z direction are not unique. Both processes simply take
care of the enhanced propagation in the z-direction, thus reducing
drastically the acceptance of charged particles from DMA in the halo.
As a result, the statement that the DMA interpretation of the EGRET is
"excluded by a large margin" because of the overproduction of
antiprotons, as claimed by \cite{bergstrom1} is only valid within a
propagation model based on isotropic propagation. Models with different
pro\-pagation in the halo and the disk can perfectly describe all observations
including DMA. 
\section{Conclusion}\label{conclusion}
Tracing of charged particles in realistic models of the regular Galactic
magnetic fields with a turbulent (small-scale) component has shown that CRs
remember the regular field lines, even if the irregular component is of the same
order of magnitude as the regular, thus leading to enhanced diffusion in $\phi$ and $z$ (see Fig.
A1 in \cite{blasi}).  With such an anisotropic  propagation model the amount of
antiprotons expected from DMA can be reduced by one to two orders of
magnitude. Therefore the claim by \cite{bergstrom1} that the DMA interpretation of
the EGRET excess of diffuse Galactic gamma rays is excluded is strongly
pro\-pagation model dependent. It only applies to a pro\-pagation
model with isotropic diffusion. An
anisotropic pro\-pagation model with different pro\-pagation in the halo and
the disk can reconcile the EGRET excess with the antiproton flux and the ratios of secondary/primary and unstable/stable
nuclei. Clearly the DMA search for light DM particles is propagation
model dependend. 

Taking these uncertainties into account shows that DMA is a viable explanation of the EGRET excess of diffuse
Galactic gamma rays, as shown in \cite{us} and and can by no means be excluded by the
antiproton flux predicted by a specific model.

\end{document}